\documentstyle[pre,aps,preprint,epsfig]{revtex}
\begin{document} 
\draft
\title{The adsorption-desorption model and its application to 
vibrated granular materials}
\author{J. Talbot$^1$, G. Tarjus$^2$, and P. Viot$^{2,3}$} \address{$^1$
Department  of   Chemistry and    Biochemistry,  Duquesne  University,
Pittsburgh,  PA 15282-1530\\   $^2$Laboratoire de  Physique
Th{\'e}orique  des Liquides,  Universit{\'e}  Pierre et  Marie Curie, 4, place
Jussieu  75252 Paris,  Cedex,  05  France\\$^3$ Laboratoire de
Physique  Th{\'e}orique, Bat.  211,   Universit{\'e} Paris-Sud 91405  ORSAY
Cedex France }
\maketitle
\begin{abstract}
We  investigate  both analytically  and  by  numerical simulation  the
kinetics of  a microscopic  model of hard  rods adsorbing on  a linear
substrate,  a  model which  is  relevant  for  compaction of  granular
materials.   The computer  simulations use  an  event-driven algorithm
which is particularly efficient at  very long times.  For a small, but
finite desorption  rate, the system reaches an  equilibrium state very
slowly, and  the long-time kinetics display  three successive regimes:
an algebraic one where the  density varies as $1/t$, a logarithmic one
where  the  density  varies  as  $1/\ln(t)$, followed  by  a  terminal
exponential approach.  The characteristic relaxation time of the final
regime, though incorrectly predicted by a mean field arguments, can be
obtained  with a  systematic gap-distribution  approach.   The density
fluctuations at equilibrium are  also investigated, and the associated
time-dependent  correlation  function  exhibits  a  power  law  regime
followed by a  final exponential decay.  Finally, we  show that denser
particle  packings can  be  obtained by  varying  the desorption  rate
during the process.
\end{abstract}

\section{Introduction}
The packing of granular   materials is somewhat paradoxical.   A child
learns quickly that it is necessary to shake a bucket in order to pack
the sand  inside,     but  physicists cannot     provide  a completely
satisfactory explanation of the densification process.  The absence of
a reference model, like the hard-sphere fluid for liquid-state physics
or  Ising model for phase transitions  and magnetism, is at the origin
of the slow progress in  this field, despite a  renewal of interest in
recent years\cite{JN96}.

To  capture   the  main  features   of  the  packing   mechanism,  the
experimental study of a model system as simple as possible can help in
building   a   reference   theory.    In  this   spirit,   Knight   et
al\cite{KFLJN95}  have considered a  system of  monodisperse spherical
beads.  The packing  process is realized by placing  beads in a column
which  is tapped  periodically with  a  given intensity.   In a  first
series of  experiments, they  demonstrated that the  density increases
monotonically  with the  number  of taps  for  various intensities  of
tapping.   The very  slow  increase  of density  was  analyzed, and  a
formula  expressing  the  density  in  terms of  the  inverse  of  the
logarithm of the number of taps was shown to be more accurate than any
of the  other suggestions\cite{KFLJN95}.   Such behavior is  common to
models  whose  geometric exclusion  effects  dictate  the kinetics  of
densification,  i.e., models  in  which addition  of  new particles  is
exponentially    limited    by     the    inverse    of    the    free
volume\cite{BG97,EG99,CLHN97,L96,PO96}.

In a second series  of experiments, Nowak et al \cite{NKPJN97,NKBJN98}
showed the presence of reversible/irreversible cycles. 
The beads in an initially loosely compacted state were vibrated for fixed
periods with a sequence of increasing vibrational intensity, causing the
density to increase monotonically. The sequence was then reversed so that
the powder was vibrated with decreasing intensity. The density, however,
continued to {\it increase} showing that the initial branch is
irreversible. When the initial sequence of increasing vibration was
repeated, the second branch was retraced confirming that it is reversible.

In the  same experiments, Nowak et  al\cite{NKPJN97,NKBJN98} monitored
the power spectrum of the density fluctuations around the steady state
for different values of the tapping strength.  A two-step spectrum was
observed  characterized  by two frequencies   that  both increase with
increasing tapping strength.    To account for  the slow  kinetics  of
compactification, the  existence of a ``reversible'' steady-state, and
the     fluctuation   power   spectrum,   they   proposed   a   simple
adsorption-desorption  or ``parking lot''model.   Partial  analyses of
this  model have   already  been reported\cite{JTT94,KB94,BKNJN98}  We
present here a comprehensive  description of the kinetics, including the
final exponential  regime and of  the fluctuations around the steady state
(equilibrium).  We first present  the model in  section II.  We detail
in  section III the  specific  algorithm  that  we have developed   for
enhancing the frequency  of  rare events  in  the late stages  of  the
densification process.  In  section   IV we study   the  densification
kinetics. By using a gap distribution analysis we derive an expression
for the  time of relaxation towards equilibrium  and the form  of the gap
distribution function in  the limit of  small desorption; the  results
compare very well with the  simulation data.  A  short account of this
derivation has  been given  in   Ref\cite{TTV99}. In section  V,   the
time-dependent density-density correlation function  is studied in the
equilibrium    state.   The correlation   function displays  two well
separated timescales, corresponding to  two relaxation steps, and this
can be interpreted by  a simple model.  In section  VI, we show that a
faster  densification can be obtained by  changing the adsorption rate
during the process.

\section{The Model}
In  the   adsorption-desorption  model, particles   are  placed  in  a
$D$-dimensional space at randomly  selected positions with  a constant
rate $k_{+}$.   If the trial particle does  not overlap any previously
adsorbed particle,  the new particle  is  accepted.  In  addition, all
adsorbed particles are subject  to removal (desorption) at random with
a constant rate $k_{-}$.  The one-dimensional version of the model, in
which the substrate is a line and the objects are  hard rods, has been
solved  in  some limiting  cases.   When $k_{-}=0$,  the adsorption is
totally irreversible and   the  process corresponds  to a  1D   Random
Sequential Adsorption   (RSA)   for  which the   kinetics   are  known
exactly\cite{E93}.   Without a relaxation  mechanism,  this process is
driven towards a non-equilibrium state and  the long-time kinetics are
given by an     algebraic scaling law, $\rho_{\infty}-\rho(t)\sim    1/t$,  with
$\rho_\infty\simeq 0.747..$ (when the substrate is empty at the beginning of the
process).   When  $k_{+}=0$,   starting   with any    configuration of
particles,  one  obtains an  analytical    solution for  this  uniform
desorption process\cite{VTVT97}.  The limit $k_{-}\to0^+$, which allows
a small but non-zero possibility of rearrangement  of the particles on
the line, leads to a final  density equal to $1$.   It is worth noting
the  finite  discontinuity  between the   final  density  of this case
($k_{-}\to0^+$)  and the RSA  jamming limit ($k_{-}=0$).  Moreover, the
final density is independent of the initial configuration of particles
on the line, whereas  the jamming limit   for the RSA process  depends
strongly  on  the  initial  state  of  the  line.  For  $ k_{-}\to0^+$,
accurate  descriptions  have been  obtained\cite{JTT94,KB94}.  In this
case, the process cleanly divides into two sub-processes.  The initial
phase consists of an irreversible adsorption and  it is followed by an
infinite  sequence of  desorption-adsorption   events in which   a rod
detaches from the surface and  the gap that  is created is immediately
filled by one or   two new rods.   The  latter possibility  causes the
system to evolve continuously to the close-packed state with $\rho=1$ as
\cite{JTT94,KB94} $  1-\rho(t) \simeq 1/\ln(t) $  where $t$ now represents a
rescaled time.   For the general case,  where both $k_+$ and $k_-$ are
non zero, a complete solution is not available.

The properties  of the adsorption-desorption model depend  only on the
ratio $K =  k_{+}/k_{-}$.  A large value of $K$  then corresponds to a
small desorption rate.  If time is expressed in units of $k^{-1}_{+}$,
the densification kinetics is given by
\begin{equation}
\frac{d\rho}{dt} = \Phi(t)-\frac{\rho}{K},
\label{kinet}
\end{equation}
where  $\Phi(t)$, the insertion  probability, is  the fraction  of the
substrate that is available for  the insertion of a new particle.  The
presence  of a  relaxation mechanism,  i.e., competing  desorption and
adsorption with  an equilibrium constant $K$, implies  that the system
eventually reaches  a steady state that corresponds  to an equilibrium
configuration  of  hard particles  with  $\rho_{\rm  eq} =  K\Phi_{\rm
eq}(\rho_{\rm  eq})$, where  $\rho_{\rm eq}$  denotes  the equilibrium
density.  At  equilibrium, the insertion probability  is given exactly
by
\begin{equation}
\Phi_{\rm eq}(\rho)= (1-\rho)\exp(-\rho/(1-\rho)).
\label{phieq}
\end{equation}
Inserting   Eq.~(\ref{phieq})  in   Eq.~(\ref{kinet})  leads   to  the
following expression for the equilibrium density:
\begin{equation}
\rho_{\rm eq} = \frac{L_w(K)}{1+L_w(K)}
\label{isotherm}
\end{equation}
where $L_w(x)$  (the Lambert-W function) is the  solution of $x=ye^y$.
In  the limit  of small  $K$, the  isotherm takes  the  Langmuir form,
$\rho_{eq}\sim   K/(1+K)$,  while   for   large  $K$,   $\rho_{eq}\sim
1-1/\ln(K)$.  At small values of $K$, equilibrium is rapidly obtained,
but at large values the densification process is dramatically slowed.

\section{Simulation algorithm} 

A naive  method of simulating the  adsorption-desorption process would
attempt to  randomly insert  a new particle  at fixed  time intervals.
This  approach, however,  is extremely  inefficient at  high densities
since most  attempts to  add new particles  are unsuccessful  after an
initial period.  We have developed a general algorithm that enables us
to investigate  in detail  the kinetics of  the adsorption  process at
arbitrarily  long  times  and  for  arbitrarily large  values  of  the
equilibrium constant,  $K$.  Instead of  using a fixed time  step, the
algorithm is based on adsorption  or desorption {\em events}.  In this
section we describe  the general features of the  algorithm that could
be to simulate an  adsorption-desorption process of arbitrarily shaped
particles in any  dimension. Later, we detail the  methodology for the
hard-rod system.

The total rate of adsorption and desorption events is
\begin{equation}
R_{\rm tot}(t) = \Phi(t) + \rho(t)/K.
\end{equation}
The quiesence, or  waiting time, is the time  interval between any two
successive events that  alter the state of the  system.  Let $F(\tau)$
denote  the  probability  that   the  waiting  time  is  greater  than
$\tau$. Since successive events are considered to be independent,
\begin{equation}
F(\tau+\Delta\tau)       =       F(\tau)(1-R_{\rm       tot}\Delta\tau
+O(\Delta\tau^2)).
\end{equation}
Taking the limit $\Delta\tau\to 0$ we obtain
\begin{equation}
F(\tau)=\exp(-R_{\rm tot}\tau).
\end{equation}
A  uniformly distributed random  number, $0<\xi_1<1$,  may be  used to
sample a random waiting time consistent with this distribution:
\begin{equation}
\tau = -\ln(\xi_1)/R_{\rm tot}.
\label{wait}
\end{equation} 
Once the quiesence time has been  selected, the nature of the event is
determined stochastically by defining
\begin{equation}
r_d(t) = \frac{\rho(t)/K}{R_{\rm tot}(t)}
\label{type}
\end{equation}
and choosing  a second  uniformly distributed random  number, $0<\xi_2
<1$. If  $\xi_2<r_d(t)$ the event  at time $t$  is a desorption  and a
randomly selected  particle is removed  from the system.  If  $\xi_2 >
r_d(t)$  the event  is  an adsorption  and  a new  particle is  placed
randomly in  the available  surface. We have  validated the  method by
applying it  to the Langmuir  equation where $\Phi  = 1-\rho/\rho_{\rm
max}$ and  the kinetics, as well  as the isotherm,  are known exactly.
This  method   is  quite  general  and   can  apply  to   a  range  of
adsorption-desorption processes.

In the simulation, the initial state  of the system  is an interval of
length $L$ (measured  in rod lengths)  bounded  by two immovable  rods
centered at   positions  $x_0=-1,   x_{N+1}=L+1$.   For  an  arbitrary
configuration of  $N$ additional rods,  whose  centers are  located at
$\{x_i, i=1,\ldots N\}$, the total available length is known exactly: $L_0
=   \sum_{i=0}^{N+1}\max(x_{i+1}-x_{i}-2,0)$.   At  each  step  of   the
simulation, the total rate of   adsorption  and desorption events   is
determined  from $R=L_0+N/K$.   A   waiting time is  sampled from  the
exponential waiting  time  distribution using Eq.(\ref{wait})  and the
type of event  is decided  with  Eq.  (\ref{type}).  If the  event  is
adsorption, a  new particle  is placed  in the available  length.  The
probability that a    particular  gap is   occupied is   equal  to its
available  length divided by the  total available  length$L_0$. Thus a
random number $\xi_3$ is generated and  the position of the particle on
the available length  is $\xi_3L_0$, which  means that  the gap between
particles $j$   and $j+1$ is  occupied   where $j$ is  defined by  the
following equation:
\begin{equation}
\sum_{i=0}^{j-1}\max(x_{i+1}-x_i-2,0)      <      \xi_3     L_0      <
\sum_{i=0}^{j}\max(x_{i+1}-x_i-2,0).
\end{equation}
Note that the adsorption event  is uniform and is always accepted.  If
the event is  desorption a particle is selected  at random and removed
from the surface. Note  that the desorption probability is independent
of the length of time that  the particle has been on the surface.  The
available line  is updated; it always  decreases (increases) following
an  adsorption  (desorption)  event.   The simulation  procedure  thus
generates    a   sequence    of    configurations   $(t_1,N_1,L_{01}),
(t_2,N_2,L_{02}),...$  and one  knows the  state of  the system  at an
arbitrary time, $t$.  To  insure good statistics, several thousands of
independent simulations must  be run for each value  of the desorption
rate. We used system lengths, $L$, from $400$ to $5000$.

\section{Gap density approach}

The  adsorption-desorption   model can be   described  in terms of gap
distribution functions.  The one-gap distribution  function, $G(h,t)$,
represents the density  of voids of length  $h$; the time evolution of
$G(h,t)$ is given by
\begin{eqnarray}
\frac{\partial{G(h,t)}}{\partial   t}&=&  -H(h-1)(h-1)   G(h,t)   +  2
\int_{h+1}^{\infty}    dh'   G(h',t)\nonumber\\   &&-\frac{2}{K}G(h,t)
+\frac{H(h-1)}{K\rho(t)}               \int^{h-1}_{0}              dh'
G(h',h-1-h',t),\label{eqcin1}
\end{eqnarray}
where $G(h,h',t)$ is the two-gap distribution function associated with
the probability  of finding two  {\it neighboring} gaps  (separated by
one particle) of length $h$ and $h'$, and $H(x)$ is the Heaviside step
function ($H(x)=1$ for $x>1,  H(x)=0$ otherwise).  The first two terms
on the  right-hand side of  Eq.~(\ref{eqcin1}) correspond to  loss and
gain  terms due  to  adsorption while  the  remaining two  are due  to
desorption.  Similarly, the time evolution of $G(h,h',t)$ is given by
\begin{eqnarray}
\frac{\partial{G(h,h',t)}}{\partial t}&=& -(H(h-1)(h-1)+H(h'-1)(h'-1))
 G(h,h',t)\nonumber\\   &+&   \int_{h+1}^{\infty}  dh''   G(h'',h',t)+
 \int_{h'+1}^{\infty}    dh''    G(h,h'',t)+    G(h+h'+1,t)\nonumber\\
 &-&\frac{3}{K}G(h,h',t)  +\frac{H(h-1)}{K\rho(t)} \int^{h-1}_{0} dh''
 G(h'',h-1-h'',h',t)\nonumber\\      &      +&\frac{H(h'-1)}{K\rho(t)}
 \int^{h'-1}_{0} dh'' G(h,h'-1-h'',h'',t),
\label{eqcin2}
\end{eqnarray}
where  $G(h,h',h'',t)$ is  the three-gap  distribution  function.  The
kinetics  of the process  is thus  given by  an infinite  hierarchy of
equations  involving   an  infinite  set   of  multi-gap  distribution
functions.

The quantities of  interest can be expressed in  terms of integrals of
the  one-gap  distribution  function.  In  particular,  the  insertion
probability $\Phi(t)$ is given by
\begin{equation}\label{phi}
\Phi(t)=\int_1^\infty dh (h-1) G(h,t),
\end{equation}
and we have the following sum rules:
\begin{equation}\label{rho1}
\rho (t)=\int_0^\infty dh G(h,t),
\end{equation}
and
\begin{equation}\label{rho2}
1-\rho (t)= \int_0^\infty dh\,h G(h,t).
\end{equation}
(One  also has  $G(h,t)=\int_0^\infty dh'G(h,h',t)=\int_0^\infty dh'G(h',h,t)$ and
similar integrals  for higher-order terms).  The steady-state solution
of Eq.~(\ref{eqcin1})   is known and  corresponds  to  the equilibrium
hard-rod system with a gap distribution function given by
\cite{JTT94,KB94};
\begin{equation}
G_{\rm        eq}(h,\rho)        =       \frac{\rho^2}        {1-\rho}
\exp\biggr(-\frac{\rho}{1-\rho}h\biggr),
\label{geq}
\end{equation} 
and    all   higher-order  distribution   functions    satisfying  the
factorization property,
\begin{equation}
G_{\rm eq}(h_1,h_2,\cdots,h_n,\rho)= G_{\rm eq}(h_1,\rho)G_{\rm eq}(h_2,\rho) \cdots
G_{\rm eq}(h_n,\rho).
\label{factor}
\end{equation}

In order to find a solution for the kinetics  of the process, one must
truncate the  hierarchy by  means of a  closure ansatz.   The simplest
closure is provided by by  an adiabatic (mean-field) treatment. There,
one assumes   that,  at any  density  $\rho(t)$,  the structure  of  the
adsorbate, as characterized by the gap distribution functions, is that
of an equilibrium system at density $\rho(t)$.

\section{Densification kinetics (at constant $K > > 1$)}
\subsection{A succession of regimes}
\begin{figure}
\begin{center}
\resizebox{12cm}{!}{\includegraphics{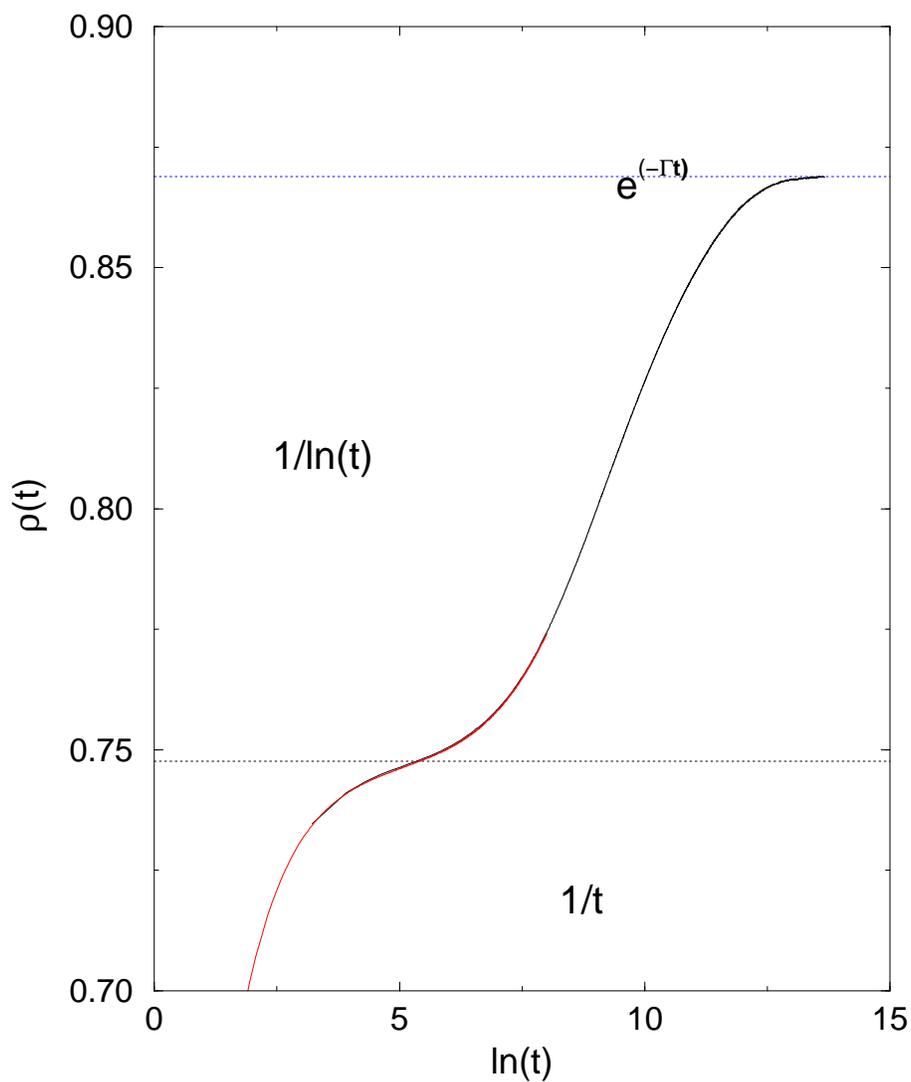}}
\caption{Linear-logarithmic plot  of the adsorbed density as a  function of time  for a
large value  of $K$ (K=5000).   The process  is characterized by three
slow kinetic   regimes:  (i)  RSA-like regime    whose  final stage is
described by a  $1/t$  behavior, (ii) $1/   \ln(t)$ regime, and  (iii)
exponential approach towards equilibrium.}\label{fig:1}
\end{center}
\end{figure}

We focus here  on  the small  desorption limit ($K  >  > 1$).   For an
initially empty line, there are   three different successive   kinetic
regimes. The  first stage  is dominated  by adsorption events  and the
process displays an RSA-like  behavior, characterized by a $1/t$ power
law  dependence.   For  $\rho(t)\geq0.7$, adsorption  becomes  slower  and
desorption     can  no longer be     ignored,   which allows  particle
rearrangements on the line and,   eventually, insertion of  additional
particles.  The densification mechanism  requires the rearrangement of
an increasing number of particles in order to open a hole large enough
for the  insertion an additional  particle.  The process is similar to
what occurs in the limit $k\to 0^+$, and the kinetics is dominated by a
$1/\ln(t)$ behavior \cite{JTT94,KB94}.  For large but finite values of
$K$, this  densification  regime continues until  the  density is very
close to the equilibrium  (steady-state) value, $\rho_{\rm eq}(K)$.   In
the  final  regime, the   desorption  term becomes  comparable  to the
adsorption term, and  an   exponential  approach to   equilibrium   is
observed.   Figure    \ref{fig:1} illustrates  the   three  successive
regimes.

\subsection{Exponential approach to equilibrium}

\begin{figure}
\begin{center}
\resizebox{14cm}{!}{\includegraphics{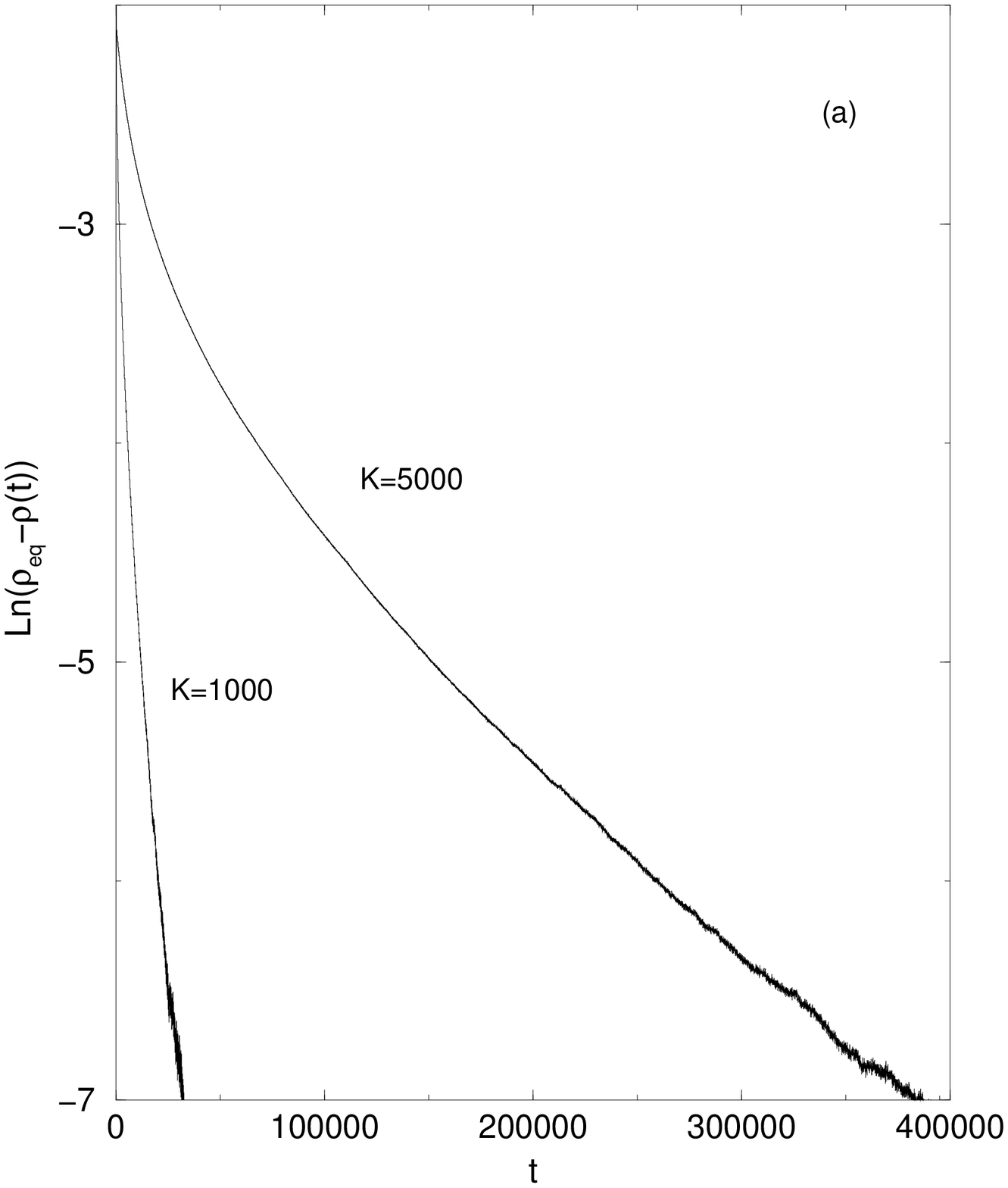}\includegraphics{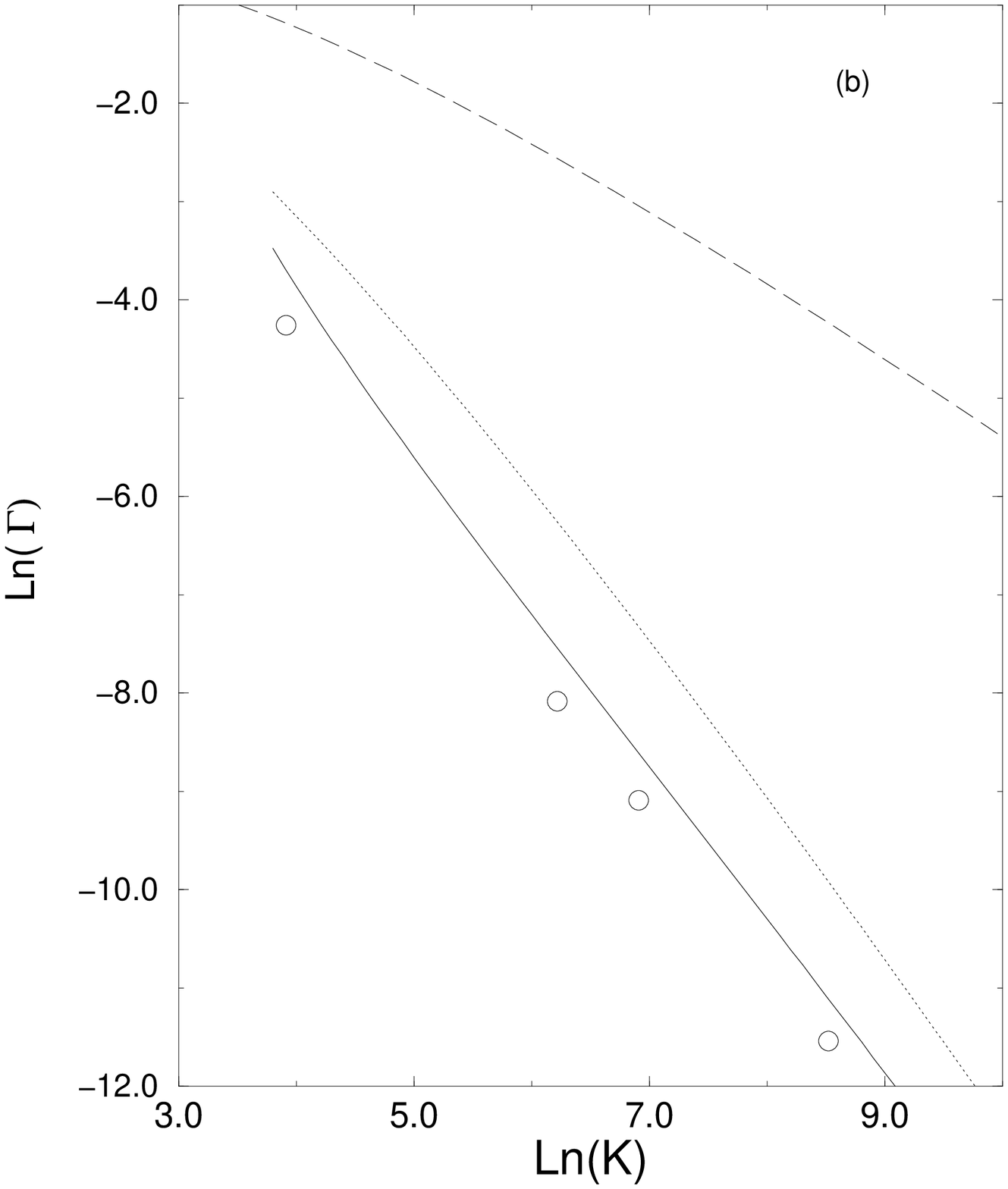}}

\caption{(a)  Final  exponential  approach   of  the  density $\rho$ to  its
equilibrium value  $\rho_{eq}$  for two large values   of  $K$.  (b)
Relaxation  rate  for the  approach to  equilibrium  $\Gamma$ versus  $K$.
Upper    curve:      prediction   from    mean-field    approximation,
Eq.~(\ref{aeq}).  Dotted  curve: leading tem  of Eq.~(\ref{aex}). Full
curve, Eq.~(\ref{eq:26}) .  Open circles: best exponential fit to the
numerical simulations.}\label{fig:2}
\end{center}
\end{figure}

The exponential regime is illustrated in  Figure ~\ref{fig:2}a.  In an
adiabatic  (mean-field)  treatment the insertion probability, $\Phi(t)$,
satisfies an  equation similar  to  Eq.~(\ref{phieq}) with $\rho(t)$  in
place of  $\rho_{eq}$.   Denoting $\delta  \rho(t) =  \rho(t)   - \rho_{\infty}$, with
$\rho_{\infty}=  \rho_{eq}(K)$, one  obtains from  Eq.~(\ref{kinet}), at first
order in density,
\begin{equation}
\frac{d}{dt}\delta\rho = -\Gamma_{MF}(K)\delta\rho + O(\delta\rho^2)
\label{expand}
\end{equation}
with
\begin{eqnarray}
\Gamma_{MF}(K)  &=&  \frac{(1+L_w(K))^2}{K},\\  &\simeq  &  \ln(K)^2/K
\,\,\mbox{\rm when K is large},
\label{aeq}
\end{eqnarray}
which is  equivalent to  a relaxation time  given by  $K/\ln(K)^2$ for
large $K$.

\begin{figure}
\begin{center}
\resizebox{12cm}{!}{\includegraphics{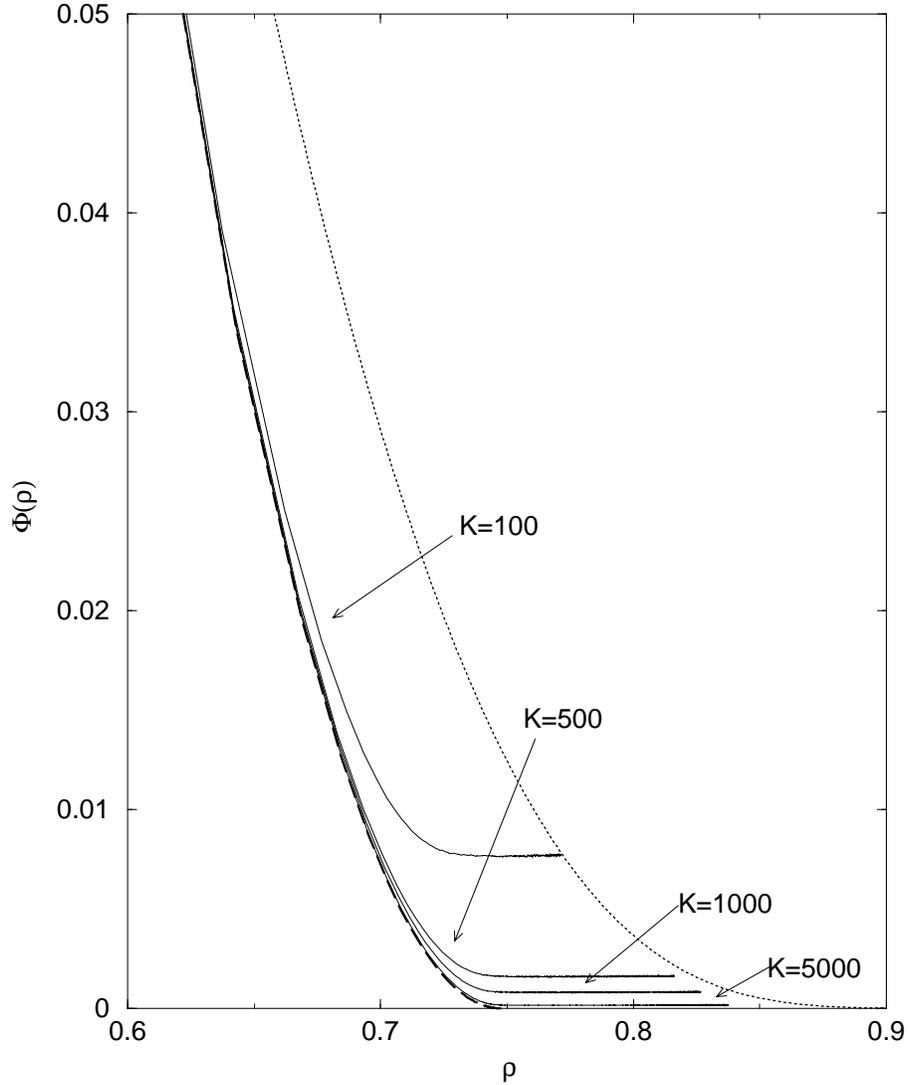}}
\caption{The  insertion probability  $\Phi$ as  a function  of density
$\rho$  for various large  values  of $K$ ($K=100,500,1000,5000$).   The
dashed curve corresponds to a process without desorption (RSA process)
and   the dotted  curve    corresponds to  the   equilibrium insertion
probability, Eq.~(\ref{phieq}).}\label{fig:3}
\end{center}
\end{figure}

In Fig.~\ref{fig:2}b, the relaxation rate  is plotted as a function of
$K$: the dashed  curve gives the mean-field prediction, Eq.~(\ref{aeq}),
and  open  circles  correspond to  the   best  exponential fit  to the
simulation results.  It is evident  that the mean-field analysis gives
a  poor estimate of the relaxation  rate for large  $K$.  This failure
can be  understood by the noting that  the mean-field assumption leads
to  a characteristic time for  the rearrangement of  $\Phi$ of the order
$K/\ln(K)^2$, i.e., much smaller than $K$, the characteristic time for
desorption.  Since in    the  absence of surface  diffusion   process,
significant rearrangement can only occur  on  a timescale longer  than
$K$ the system is unable to  adjust rapidly enough  in order to change
significantly  the available surface  function on a timescale of order 
$K/ \ln(K)^2$.  In Fig~\ref{fig:3},we
display the insertion probability, $\Phi(\rho)$,  for several large values
of $K$; it is worth noting that $\Phi$ first follows the RSA curve until
it reaches  a value close to   the equilibrium one  at  which point it
plateaus  and evolves very   weakly towards equilibrium.  The  process
clearly deviates from the adiabatic description in which the insertion
probably  is given,  at  all  densities,  by the equilibrium   curve,
$\Phi_{\rm eq}(\rho)$.

We  now  turn  to a  description  in  terms  of the  gap  distribution
approach. To obtain the leading   term   in the exponential  approach    towards
equilibrium, when $K$ is  very large (but finite),  we assume that, as
for the steady state (or equilibrium), $|G(h,t)|\sim \exp (-\Pi h)$, with
$\Pi   \sim   \ln  K      \sim    (1-\rho)^{-1}$  (see    Eqs.~(\ref{geq}) and
(\ref{isotherm}),  when $K$ is very large).   As a consequence, if one
defines              $\rho_n(t)=\int_n^{n+1}G(h,t)dh$                  and
$\Phi_n(t)=\int_n^{n+1}(h-1)G(h,t)dh$, then $\rho_n\sim \Phi_{n}\sim K^{-n}$, so
that if one looks for the dominant behavior in $1/K$, it is sufficient
to  consider the   first  intervals  in  $h$.    As in  the  adiabatic
approximation, one can expand the gap  densities in power of $\delta\rho(t)$
and  keep only  the linear term  which  gives rise  to the exponential
decay.  In the final  regime, where  the density  is close  to the
steady  state,  we  first  assume   that  the  deviation  of  the  gap
distribution   function    from   its   equilibrium    form,   $\delta
G(h,t)=G(h,t)-G_{eq}(h)$,  can   be  expressed  as   an  expansion  in
$\delta\rho(t)$ where only the first term is kept.  Let us then denote
\begin{equation}\label{defah}
A(h)=\rho_{\infty}\left.\frac{\partial                              \ln
G(h,\rho)}{\partial\rho}\right|_{\rho_\infty}
\end{equation}
and
\begin{equation}\label{defa}
A(h,h')=\rho_{\infty}\left.\frac{\partial  \ln  G(h,h',\rho)}{\partial
\rho}\right|_{\rho_\infty}.
\end{equation}
After inserting Eq.~(\ref{defah})  in Eq.~(\ref{eqcin1}) $A(h)$ can be
rewritten for $0<h\leq 1$ as
\begin{equation}\label{defgam}
\frac{2-\gamma}{K}A(h)=\frac{2P_\infty}{K}
\int_h^{+\infty}dh'e^{-P_\infty(h'-h)}A(1+h'),
\end{equation}
where  $P_\infty=\rho_\infty/(1-\rho_\infty)$   is  the  dimensionless
equilibrium pressure for $\rho_\infty = \rho_{eq}$ and
\begin{equation}\label{eq:1}
\gamma= K\Gamma= -K\delta\dot{\rho(t)}/ \delta\rho(t)|_{\rho_\infty}.
\end{equation}
From Eqs.~(\ref{kinet}) and~(\ref{phi}), one obtains
\begin{equation}\label{gameq}
\gamma=1 -P^2_\infty\int_0^\infty dh h e^{-P_\infty h} A(1+h).
\end{equation}
whereas  the sum  rules  in Eq.~(\ref{rho1})  and ~(\ref{rho2})  give,
respectively,
\begin{equation}\label{rule1}
P_\infty\int_0^{\infty} dh e^{-P_\infty h}A(h) = 1
\end{equation}
\begin{equation}\label{rule2}
  -P_\infty \int_0^{\infty} dh h e^{-P_\infty h} A(h)=1.
\end{equation}
When integrating the two  sides of Eqs.~(\ref{defgam}) between $0$ and
$1$, one obtains
\begin{equation}\label{eq:2}
(2-\gamma)=2P^2_\infty\int_0^1dh h e^{-P_\infty h}A(1+h)+O(1/K).
\end{equation}
Combining     Eq.~(\ref{gameq})    with     Eq.~(\ref{eq:2})    yields
$\gamma=O(1/K)$.  Thus, the  relaxation rate $\Gamma$ goes essentially
as $1/K^2$  instead of  the $1/K$ dominant  behavior predicted  by the
mean-field treatment.  In order to have a more explicit expression for
$\gamma$ and  $\Gamma$, it is  necessary to calculate the  integral on
the right-hand side of Eq.~(\ref{gameq}) to $O(1/K^2)$:
\begin{eqnarray}\label{eq:3}
(2-\gamma)\left[1-\frac{P_\infty^2}{K}\left(\int_0^1e^{-P_\infty
h}A(1+h)+A(1)\right)\right]&=&2-2                          \left(\gamma
-\frac{P_\infty^3}{K}\int_0^1(h+1)e^{-P_\infty
h}A(2+h)\right)\nonumber \\&+& O(1/K^2),
\end{eqnarray} 
which leads to
\begin{equation}\label{eq:4}
\gamma                                   K=2P_\infty\left(A(0)+P_\infty
A(1)-P_\infty^2\int_0^1(h+1)e^{-P_\infty h}A(2+h)\right)+ O(1/K).
\end{equation} 
An explicit expression for  $\gamma K$, Eq.~(\ref{eq:4}) thus requires
the  knowledge of  the  gap distribution  function  for $3>h>1$.   The
kinetic equation for the gap  distribution function when $h>1$ is then
rewritten   by  inserting   Eqs.~(\ref{defah})  and   (\ref{defa})  in
Eq.~(\ref{eqcin1}), which gives
\begin{eqnarray}
\left(h-1+     \frac{2-\gamma}{K}\right)A(h)&=&    -(h-1)+\int_0^{h-1}
dh'A(h',h-1-h')\nonumber\\&+&\frac{2P_{\infty}}{K}\int_{h-1}^\infty dh
e^{-P_{\infty}(h'-h)}A(1+h). \label{eq:5}
\end{eqnarray}
Combining Eq.~(\ref{eq:4}) and Eq.~(\ref{eq:5}) for $h=1$, one finally
gets
\begin{equation}\label{eq:6}
\gamma  K=2P_\infty\left(A(0)-P^2_\infty\int_0^1   dh  h  e^{-P_\infty
h}A(2+h)\right)+O(1/K).
\end{equation}
Since the system evolves close to  equilibrium, we further assume that
the factorization  property  for the two-gap distribution  function is
valid  to  $O(1/K)$, i.e.,   $A(h,h')=A(h)+A(h')+ O(1/K)$\cite{VTT93}.
Eqs.~(\ref{eq:5}) and (\ref{defgam})  then become   a closed set    of
equations for $A(h)$    to $O(1/K)$. The   solution  is  given in   the
Appendix,   as    well as  the   explicit    expression   for
$\Gamma(K)=\gamma /K$,
Eq.~(\ref{eq:26}).   As an illustration,  the leading terms of $\Gamma(K)$
in powers of $\ln(K)$ are obtained as 
\begin{equation}\label{aex}
\Gamma\simeq    2\frac{(\ln   K)^3}{K^2}  - 4\frac{(\ln K)^2}{K^2}+2\frac{(\ln
K)}{K^2}+ O(\frac{1}{K^2}).
\end{equation}
The   prediction  of Eq.~(\ref{eq:26}),   shown  as the full  curve in
Fig.~\ref{fig:2}b, gives a   good agreement with  the results obtained
from an exponential fit to the simulation data, whereas the mean-field
predictions   fail completely   (dashed  curve)   (the   dotted  curve
corresponds to the first term of the right hand-side of Eq. (\ref{aex})).

\section{Fluctuations around equilibrium}

\begin{figure}
\begin{center}
\resizebox{14cm}{!}{\includegraphics{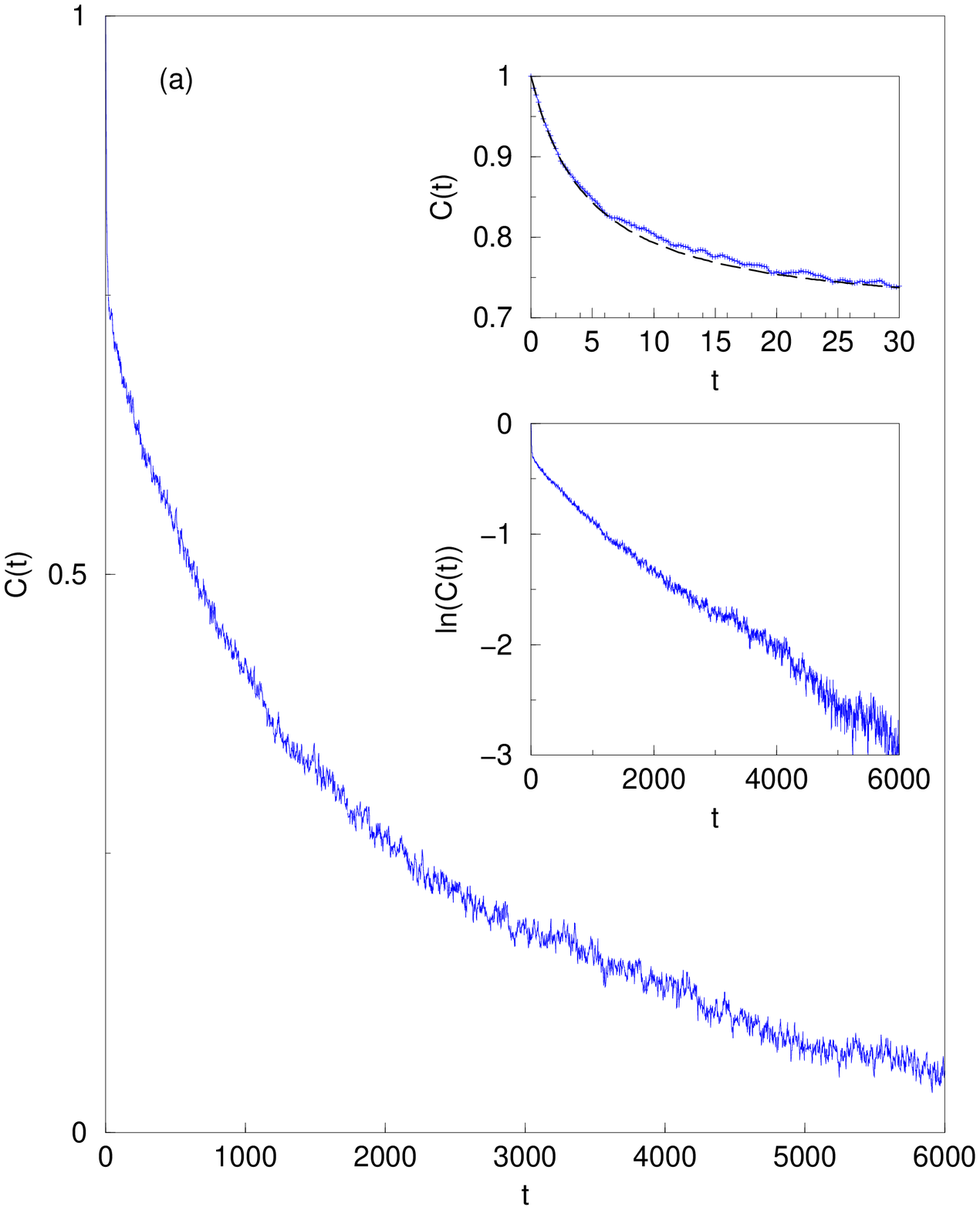}\includegraphics{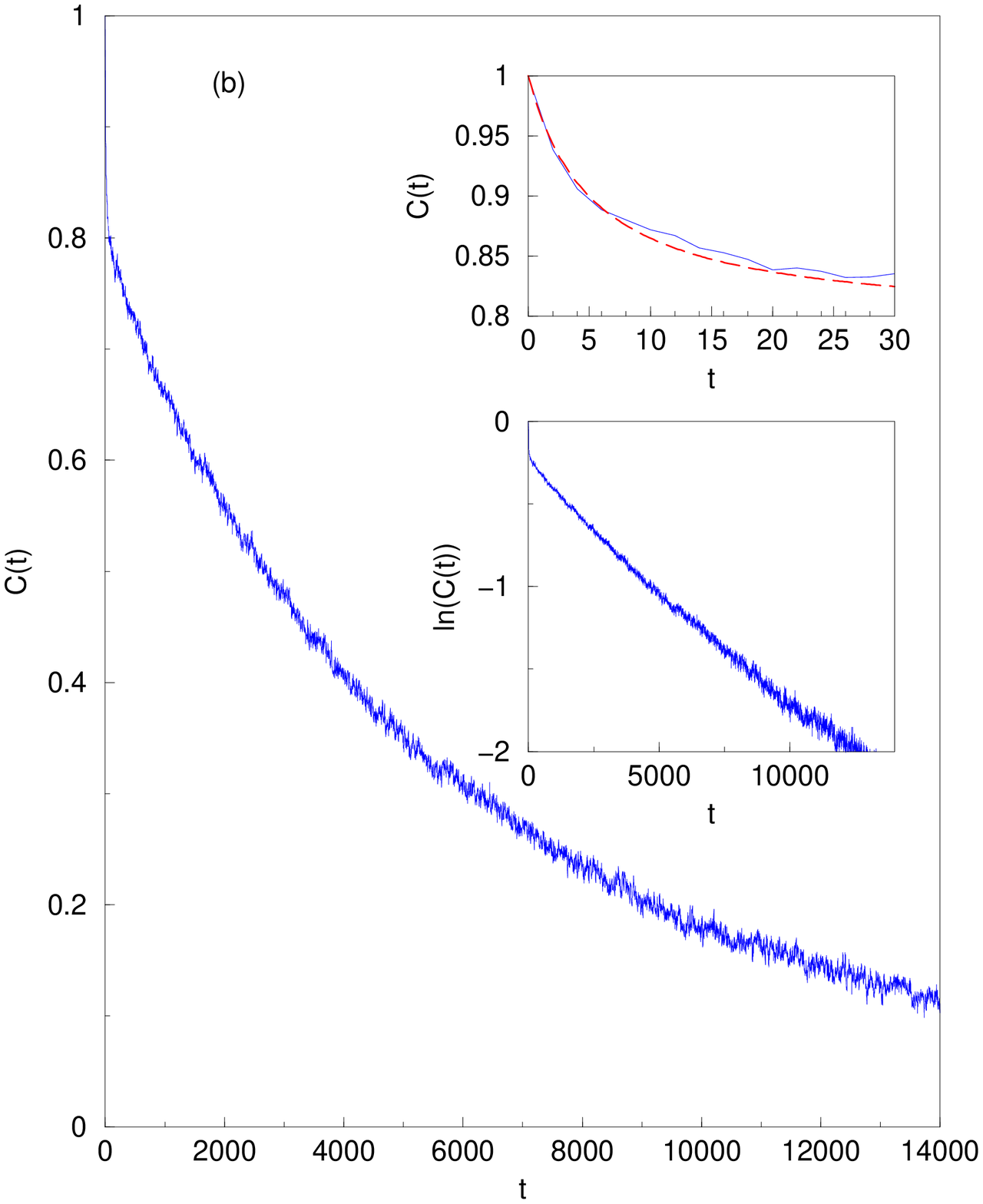}}

\caption{Equilibrium density-density correlation function $C(t)$ versus
 time for (a) $K=500$ and (b) $K= 1000$.  The inset in the upper right
 corner  displays  the first  step  in  the decay  of  the correlation
 function  (full curve) as well  as  the predicted short-time formula,
 $C_{short}(t)$, Eq.~(\ref{eq:30}) (dashed   curve).  The other  inset
 shows  the exponential decay of $C(t)$ at
 long times on a logarithmic-linear plot.}\label{fig:4}
\end{center}
\end{figure}

For times much larger than  the relaxation time, the density no longer
evolves (on average), but fluctuates around its equilibrium value.  Note
that  in this  regime,  the fluctuation-dissipation  theorem and the  time
translational   invariance  are both  valid.   We   have   calculated  the
time-dependent   correlation  function   $C(t)$   of  the   density
fluctuations,    $\delta\rho(t)=\rho(t)-\rho_{\infty}$,    around    the
equilibrium state. Starting from the $2-$time correlation function,
\begin{equation}\label{eq:7}
C(t'+t,t')=
\frac{<\rho(t+t')\rho(t')>-<\rho(t'+t)><\rho(t')>}{<\rho(t')^2>-<\rho(t')>^2},
\end{equation}
we have numerically verified  that when $t'>  > 1/ \Gamma $,  $C(t'+t,t')$
becomes time translationnally  invariant, i.e.      $C(t+t',t')=C(t)$.
(Conversely,   when  $1<   <t'< <   1/    \Gamma $,   one  observes  aging
phenomena\cite{BCKM98},  but  we   postpone the   discussion  of  this
phenomena   to a future  publication).      Because of the very   long
relaxation  time, we  found that the  calculation  of  the correlation
function  is more  efficient by   using  Eq.  (\ref{eq:7}) instead  of
taking the usual time average  on a single system\cite{AT87}.  Results
from the simulation are shown in  Fig.\ref{fig:4} for two large values
of $K$.   At  short and intermediate   times, the  decay of $C(t)$  is
non-exponential, whereas at  long times the kinetics follows an exponential
decay.   The two regimes, or  relaxation  steps, can be interpreted as
follows:    the first  consists   of  a ``fast'' adsorption-desorption
process  without appreciable densification  of the system, whereas the
second corresponds to the linear-response  regime and, as predicted by
Onsager's   regression hypothesis, it  shows the same final
exponential dependence as the final approach of $\rho(t)$ towards $\rho_\infty
$  in the densification  process.   In  this  second relaxation  step,
$C(t)\sim  e^{-\Gamma  t}$  where  $\Gamma$    is   given by    Eqs.~(\ref{aex})
and~(\ref{eq:26}).

Close to equilibrium and  for large values  of $K$, the adsorption and
desorption  events can be considered as    spatially uncorrelated, and
the system  can be represented as  a set of  independent  gaps in which a
particle is  adsorbed  or not. This  assumption does  not account  for
rearrangements which  occur   at long times,  but is   valid for short
times. When a  particle is adsorbed, the  gap is characterized by  the
distribution  $g_{eq}(h)$ which   is the   distribution probability of
finding a particle  such  the total length  of  right and left  gap is
equal to $h$, i.e.
\begin{eqnarray}
g_{eq}(h)&=&\frac{1}{\rho_\infty}\int_0^hdh' G_{eq}(h')
G_{eq}(h-h')\nonumber\\
&=&\rho_\infty P_\infty^2h\exp(-P_\infty h).
\end{eqnarray}
Once  the particle  has desorbed, the gap is characterized
by the distribution $G_{eq}(h+1)$.  The two distributions are
 calculated at equilibrium and their ratio is given exactly by:
\begin{equation}
\frac{G_{eq}(h+1)}{g_{eq}(h)}=\frac{1}{Kh}.
\end{equation} 
For a given gap of length $(h+1)$, we have a two-state (particle-hole)
stochastic  process in which the  rates, associated to   desorption and
adsorption  respectively, are  $1/K$ and $h$.  The average probability
for having a particle in the gap is equal to
\begin{equation}
P(h)=\frac{hK}{1+hK}.
\end{equation} 
For a given gap   of  length $(h+1)$, the (unnormalized)   correlation
function of the density fluctuations $\tilde{C}_h(t)$ due to the two-state
stochastic process is given by\cite{G90}
\begin{equation}
\tilde{C}_h(t)=\frac{P(h)}{(1+hK)}\exp(-(1/K+h)t).
\end{equation}
Assuming  that  the adsorption-desorption events giving rise to the two-state
process  only  seldom  affect simultaneously  two
neighboring  particles, one  can write  the   (unnormalized)
correlation function   
as a  superposition of correlation  functions occurring in  parallel in
the different gaps, weighted by the distribution $g_{eq}(h)$, i.e.,
\begin{eqnarray}
C_{short}&=&\int_0^\infty dh g_{eq}(h) \tilde{C}_h(t)\\
&=& \frac{\rho P^2_\infty}{K}\frac{\exp(-t/K)}{(t+P_\infty)}+ O(1/K^2).\label{eq:31}
\end{eqnarray}
At   equilibrium, the variance  of  the  density  fluctuations  can be
calculated exactly\cite{BBV94},
\begin{equation}
<(\delta \rho)^2>=\rho_\infty (1-\rho_\infty)^2, 
\end{equation}
so that, the normalized  correlation  function $C(t)$ at short  times,
$C_{short},$ can be written as
\begin{equation}\label{eq:30}
C_{short}(t)=\frac{ P^2_\infty}{K(1-\rho_\infty)^2}\frac{\exp(-t/K)}{(t+P_\infty)}+ O(1/K^2),
\end{equation}
which  reduces to a power  law, $1/t$, when  $\ln(K)< <t  < <K$.  This
result is equivalent   to  the $\ln(\omega  )$-behavior already  predicted
along  similar lines  similar by Kolan   {\it et al}\cite{KNT98}.  The
insets  in   Figs.~\ref{fig:4}a,b  illustrate  the excellent  agreement
between Eq.~(\ref{eq:31}) and the simulation data.

It is worth noting that the $1/t$ behavior  is reminiscent of the pure
RSA asymptotic regime, where it occurs as a consequence of the filling
of small isolated pieces of the  available fraction of the line, whose
lengths go  to zero  when  $t\to\infty$.  With  similar arguments, one thus
expects to have in higher dimensions a $t^{-1/D}$ behavior, leading to
an   $\omega^{(1-1/D)}$ power-law dependence  for  the power spectrum.  In
particular, this predicts a power law  $\omega^{-1/2}$ for $D=2$, which is
compatible    with   the  experimental   data   in  vibrated  granular
media\cite{NKBJN98}.  In   dimensions higher  than  $1$, our  prediction
differs from  that of   Kolan  {\it  et  al}\cite{KNT98} since   their
analysis leads  to a $1/ \omega   $ dependence in  the power  spectrum.  A
numerical  study    of    the    two-dimensional   version   of    the
adsorption-desorption model should settle this point.

\section{Densification regime and multistep process}

\begin{figure}
\begin{center}
\resizebox{12cm}{!}{\includegraphics{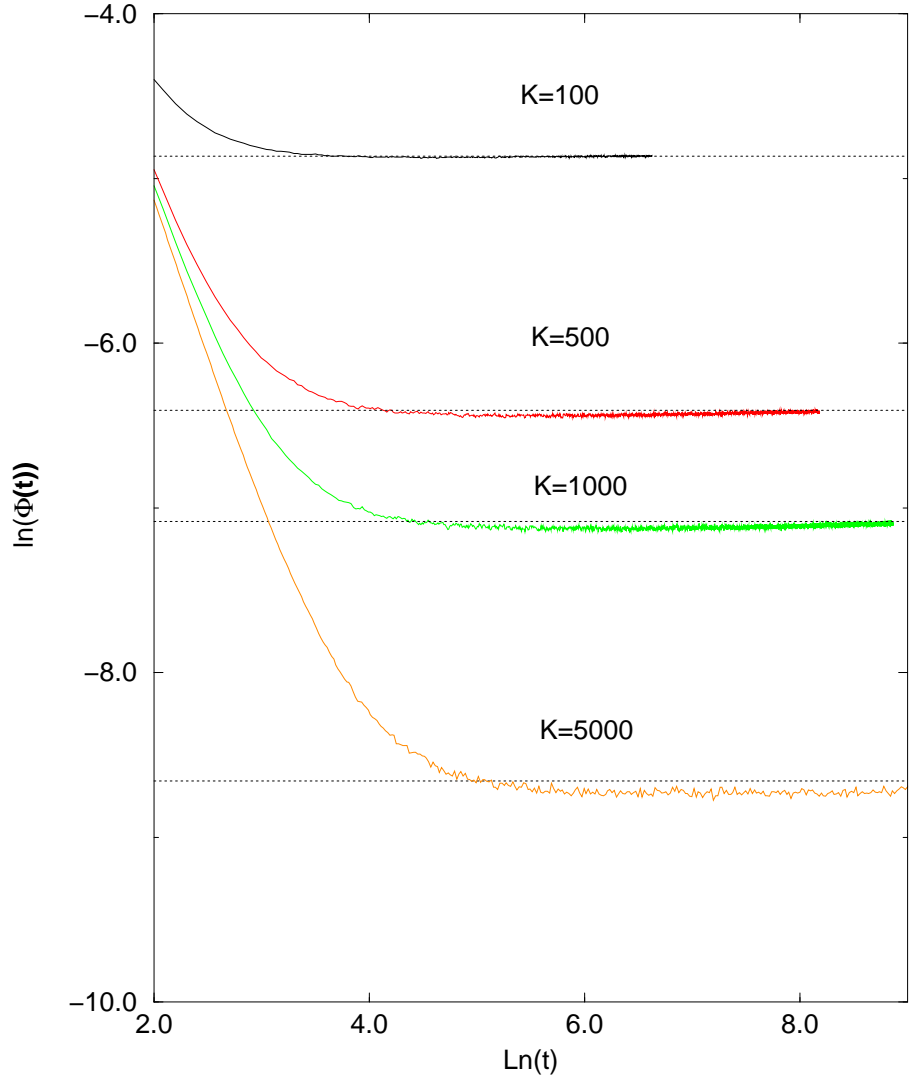}}
\caption{Log-log plot of the insertion  probability $\Phi$ as  a function of  time for
various large  values of $K$  ($K=100,500,1000,5000$).   Note that for
$K>100$, $\Phi$ displays a minimum which is smaller than the equilibrium
value. (dotted lines)}\label{fig:5}
\end{center}
\end{figure}

The    very    slow-exponential    approach  to   equilibrium     with
$\Gamma(K)=O(1/K^2)$  when  $K$   is  very large    implies  that $\Phi(\rho)$
increases with $\rho$  when $\rho$ is sufficiently  large (see  section VB
and the  Appendix).  Since in the  first  (RSA-like) regime  $\Phi (\rho)$
decreases,   there    always   exists    a   density     $\rho_m$  where
$\partial\Phi(\rho)/\partial\rho|_{\rho_m}=0$.   Fig.~\ref{fig:5} displays  a log-log plot
of $\Phi$ as a function  of time for various  values  of $K$.  One
notice (i) that $\rho_m$ is an increasing function of $K$ and (ii) that the
minimum of $\Phi$ is always very  close but smaller than the equilibrium
value,$\Phi_{\rm eq}(K)$, and  smaller than $\rho/K$, which  is due to the
fact that the density is an increasing function of time.

\begin{figure}
\begin{center}
\resizebox{12cm}{!}{\includegraphics{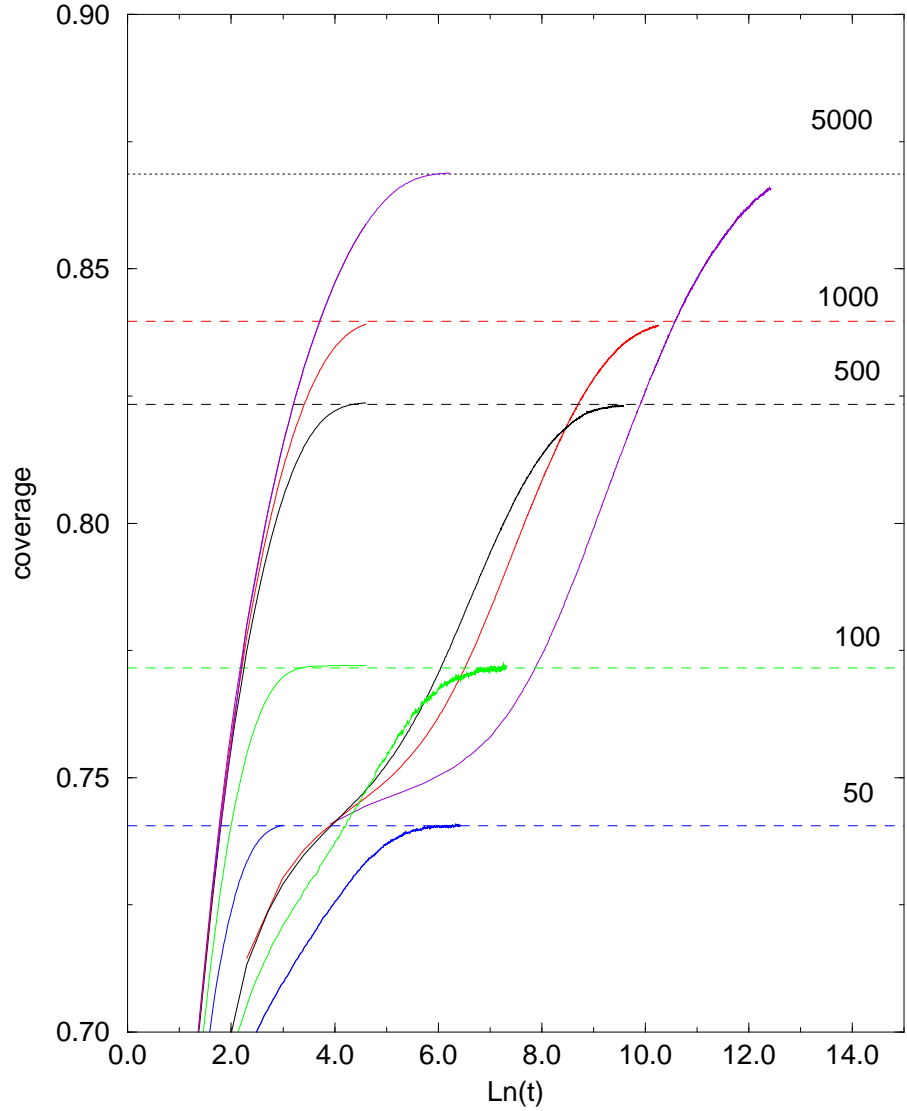}}

\caption{Linear-Log plot of the density versus time for different values of $K$.  The
left curves correspond to the adiabatic  process.  The right curves
correspond to
the adsorption-desorption model. Notice that in the latter case the
curves for different values of $K$ always cross, a phenomenon absent in the
adiabatic process}\label{fig:6}
\end{center}
\end{figure}

\begin{figure}
\begin{center}
\resizebox{12cm}{!}{\includegraphics{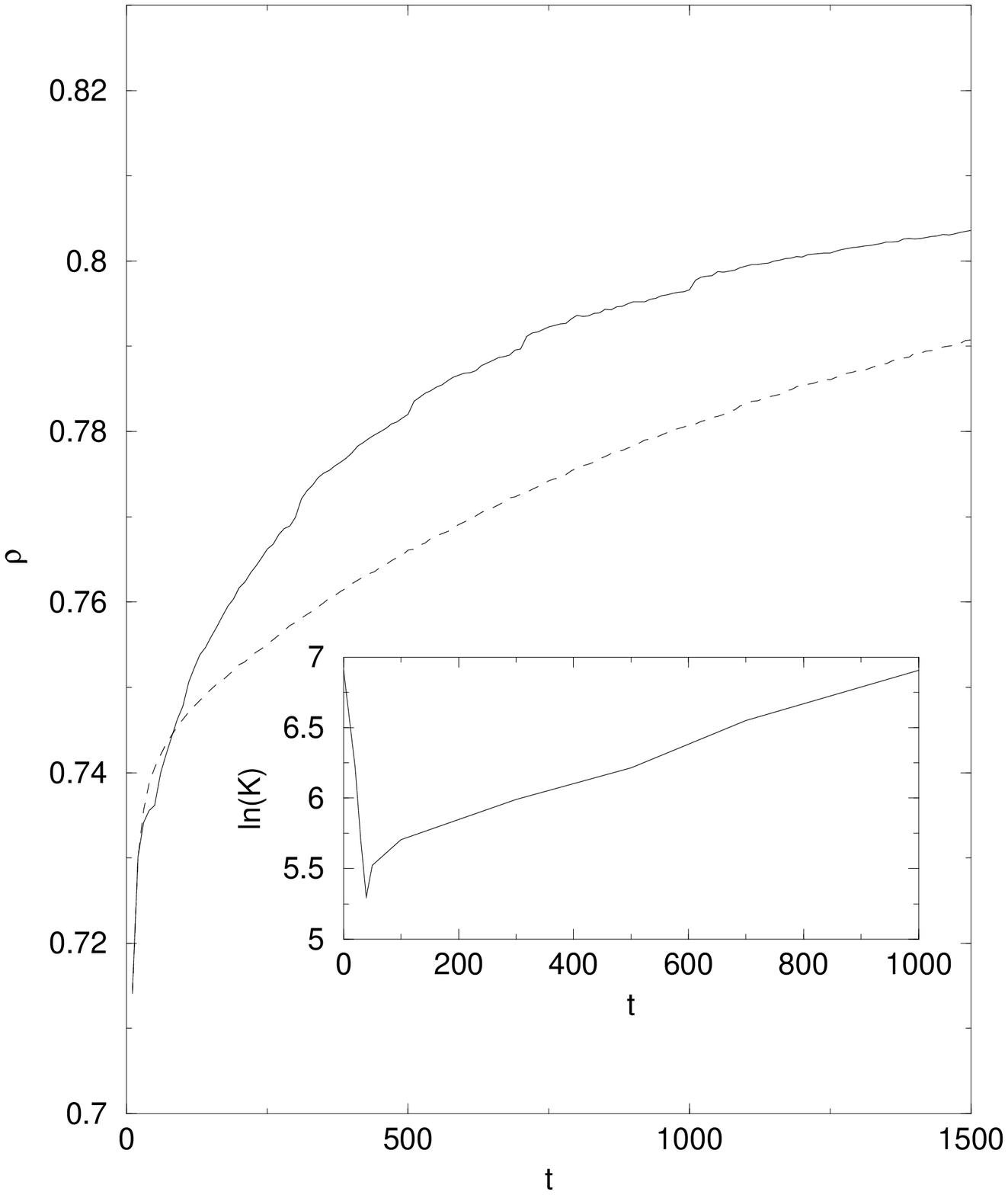}}
\caption{
Density  increase over 1500  time units  for a  process with  a single
value of K ($K=1000$) and for multistep  process in which the sequence
of  $K$  is shown on a   linear-logarithmic plot in   the inset (After
$t=1000$,  $K$ stays constant).  Note the large enhancement of packing
efficiency in the  latter  process.   This effect   is absent in   the
adiabatic  approximation  where    the density   is  a   monotonically
increasing function of $K$.  }\label{fig:7}
\end{center}
\end{figure}

In Fig.~\ref{fig:6} the  density is plotted as  a function of time for
different values  of K.  The  curves  on the  left part of  the figure
correspond   to   an adiabatic process    where the  available surface
function  is replaced by  the  equilibrium formula, Eq.~(\ref{phieq}),
and which   corresponds to a  process   where rapid diffusion   on the
substrate allows a  instantaneous equilibration after  each desorption
and adsorption event.  For all values of $K$, the adiabatic process is
much   faster than    the   corresponding adsorption-desorption  model
process.   Moreover, for an adiabatic  process, the density  is at all
times always   a monotonically increasing function  of   $K$.  For the
adsorption-desorption model, on the   other hand, the density is  {\it
not always  monotonic} in $K$.  In   Fig~\ref{fig:6}, for example, the
system with $K=500$ has a higher density than the system with $K=1000$
for $4\lesssim   \ln(t)\lesssim  8$.  The existence  of  a  minimum in $\Phi$  is  a
sufficient condition for this phenomenon.  It  follows that {\it for a
given finite time}, the  densification can be  made more  efficient by
changing  the desorption  rate during  the  process.  Fig.~\ref{fig:7}
compares  the densities  obtained by  using either  a single  value of
$K=1000$ or a sequence of varying $K$, starting  from $1000$ at $t=0$,
passing  through a minimum, and  finishing at  the same value $K=1000$
when  $t=1000$ .  One clearly observes  that a larger final density is
reached with the multistep process.  Such a  phenomenon, which is also
the source of the reversible-irreversible cycles observed by Nowak{\it
et al}\cite{NKBJN98},  has  been already observed  and  quantitatively
analyzed   in  an  irreversible   adsorption
model\cite{VTVT97}. However,  the
determination of  the   optimum   densification  strategy,  which  has
significant  applications  to    vibratory   compaction  of   granular
materials, is still an open problem.

\section*{Acknowledgements}
 
J.T.  thanks the National Science  Foundation for financial
support.  The  Laboratoire de  Physique Th{\'e}orique des  Liquides is
Unit{\'e}  Mixte de  Recherche No  7600 au  CNRS.  The  Laboratoire de
Physique  Th{\'e}orique is  Unit{\'e} Mixte  de Recherche  No  8627 au
CNRS.  \appendix
\section*{ Solution for the gap distribution function in the
linear response  region} For convenience, we  introduce the notation
$B(h)=A(h)-1$, $B(h,h')=A(h,h')-2$, etc.  For $0\leq h\leq 1$, by taking
into  account   that  $\gamma  =O(1/K)$,   Eq.~(\ref{defgam})  can  be
reexpressed as
\begin{equation}\label{eq:11}
B(h)e^{-P_\infty                     h}-B(1)\frac{P_\infty}{K}=P_\infty
\int_0^1dh'e^{-P_\infty h'}B(1+h')+ O(e^{-P_\infty h}/K)
\end{equation}
and for larger gaps, Eq.~(\ref{eq:5}) can be rewritten for $ h\geq 0$,
as                                         \begin{equation}\label{eq:12}
(h+\frac{2}{K})B(1+h)=\int_0^hdh'B(h',h-h')+\frac{2P_\infty}{K}e^{P_\infty
h}  \int_h^\infty  U_\infty  dh'e^{-P_\infty  h'}B(2+h')  +  O(1/K^2).
\end{equation} Assuming that the  factorization property is valid to a
$O(1/K)$, i.e.,
\begin{equation}\label{eq:13}
B(h,h')=B(h)+B(h')+\frac{C(h,h')}{K}+O(1/K^2),
\end{equation}
with $0\leq  h\leq 1$, and $O(1/K)$ or  $O(1/K^2)$ designate functions
that are  uniformly of order  $1/K$ or $1/K^2$ on
the interval $[0,1]$, we can derive from Eq.~(\ref{eq:12})
\begin{eqnarray}
(h+\frac{2}{K})B(1+h)-2\int_0^{h-1}dh'B(h')&=&\frac{1}{K}\left[\int_0^hdh'C(h',h-h')+
2P_\infty                 \int_h^\infty                dh'e^{-P_\infty
(h'-h)}B(2+h')\right]\nonumber\\ &+& O(1/K^2),\label{eq:14}
\end{eqnarray}
and
\begin{equation}\label{eq:15}
(1+h)B(2+h)-2\int_0^{1+h}dh'B(h')= O(1/K).
\end{equation}
When $h > > 2/K$, Eq.~(\ref{eq:14}) simplifies to
\begin{equation}\label{eq:16}
hB(1+h)-2\int_0^1dh'B(h')= O(1/K).
\end{equation}
Deriving  Eq.~(\ref{eq:16})  with respect  to  $h$  and inserting  the
result  in  Eq.~(\ref{eq:11}),  one  gets  the  following  differential
equation,
\begin{equation}\label{eq:17}
\frac{1}{P_\infty}\frac{d^2}{dh^2}(hB(1+h))-\frac{d}{dh}(hB(1+h))+2(hB(1+h))=
O(1/K)
\end{equation}
with $1\geq h>>2/K$, whose solution is
\begin{equation}\label{eq:18}
B(1+h)=b(2-P_\infty     h        )+c\left[-\frac{(1-P_\infty        h)e^{-P_\infty
(1-h)}}{P_\infty(h+2/K)}     +  (2-P_\infty   h)e^{-P_\infty}\int_0^hdh'\frac{e^{P_\infty
h'}}{h'+2/K}\right],
\end{equation}
where  $b$  and  $c$  are   constants  and  $1\geq  h>  >  2/K$.   The
corresponding solution for $B(h)$ is then
\begin{equation}\label{eq:19}
B(h)=b(1-P_\infty   h    )+c\left[e^{-P_\infty   (1-h)}+   (1-P_\infty
h)e^{-P_\infty}\int_0^hdh'\frac{e^{P_\infty h'}}{h'+2/K}\right].
\end{equation}
It is important  to stress that the above  equations give the solution
only when  $h{\geq}2/K$. To satisfy  Eqs.~(\ref{eq:11}) and (\ref{eq:13})
when $h\sim 2/K$ or smaller, on must include in $B(1+h)$ an additional
component that  is a $O(1/K)$ when $h  > > 2/K$ and  is non negligible
only when $h\sim 2/K$. The full solution for $0\leq h\leq 1$ is then obtained as

\begin{eqnarray}
B(1+h)&=&b(2-P_\infty   h  )+c\left[-\frac{(1-P_\infty  h)e^{-P_\infty
(1-h)}}{P_\infty(h+2/K)}                                   +(2-P_\infty
h)e^{-P_\infty}\int_0^hdh'\frac{e^{P_\infty
h'}}{h'+2/K}\right]\nonumber\\                   &+&\frac{d}{K(h+2/K)}+
O(1/K).\label{eq:20}
\end{eqnarray}
It is easy to verify that Eq.~(\ref{eq:19}) is still the full solution
to a  $O(1/K)$ for $0\leq  h\leq 1$ and that,  from Eq.~(\ref{eq:15}),
the solution for $B(2+h)$, $0\leq h\leq 1$, is equal to :
\begin{eqnarray}
B(2+h)&=&\frac{1}{(1+h)}\left\{B(2)+bh(4-P_\infty
h)-\frac{c}{P_\infty}\left[(3-P_\infty             h)e^{-P_\infty(1-h)}
\right.\right.\nonumber       \\      &+&(2-4P_\infty      h+(P_\infty
h)^2)\left.    \left.     e^{-P_\infty}\int_0^{h'}   \frac{e^{P_\infty
h'}}{h'+2/K}\right] + O(1/K)\right\} \label{eq:21}
\end{eqnarray}
The constants $b$, $c$, $d$,  and $B(2)$ are determined by the various
sum rules as well as by  the condition, which comes from the structure
of  the hierarchy  of kinetic  equations, that  $B(h)$ is  a piecewise
continuous function, namely,
\begin{eqnarray}\label{eq:22}
B(1)&=&2b-\frac{c}{2}+\frac{d}{2}=b(1-P_\infty)+c[1+(1-P_\infty)e^{-P_\infty}E_i(P_\infty)]+
O(1/K),\\
B(2)&=&b(2-P_\infty)+c[-\frac{1}{P_\infty}+1(2-P_\infty)e^{-P_\infty}E_i(P_\infty)]+
O(1/K),
\end{eqnarray}
where $E_i(x)=e^x/x+e^x \int_0^\infty  dt\exp(-t)/(x-t)^2$.

The result can be expressed as
\begin{eqnarray}
B(0)&=&b=P_\infty          +1\\\label{eq:23}          B(1)&=&(P_\infty
+1)\left[-(P_\infty  -1)+2(P_\infty +1) \frac{(1-P_\infty\Phi_1^\infty
)(1+(1-P_\infty)\Phi_i^\infty
)}{1+(1-P_\infty)(\Phi_1^\infty+\Phi_i^\infty)-P_\infty(2-P_\infty)\Phi_1^\infty\Phi_i^\infty}\right]\\\label{eq:24}
B(2)&=&(P_\infty  +1)\left[2-P_\infty+2\frac{(P_\infty  +1)}{P_\infty}
\frac{(1-P_\infty\Phi_1^\infty
)(P_\infty-1+P_\infty(2-P_\infty)\Phi_i^\infty)}
{1+(1-P_\infty)(\Phi_1^\infty+\Phi_i^\infty)-P_\infty(2-P_\infty)\Phi_1^\infty\Phi_i^\infty}\right],\label{eq:25}
\end{eqnarray}
where        we       have        introduced        the       notation
$\Phi_i^\infty=e^{-P_\infty}E_i(P_\infty)$                           and
$\Phi_1^\infty=e^{P_\infty}E_1(P_\infty)$  with  $E_1(x)=\int_1^\infty
dt\frac{e^{-xt}}{t}  $. The  values of  $c$ and  $d$ can  be trivially
derived from the above equations.

The relaxation  rate $\Gamma=\gamma K$ can be  obtained by inserting the  above solution
into Eq.~(\ref{eq:4}), which leads to
\begin{equation}\label{eq:26}
{\Gamma }{K^2}=\gamma K=2P_\infty(1+P_\infty)^2\frac{1-U_\infty }{1+U_\infty }
\end{equation}
with
\begin{equation}\label{eq:27}
U_\infty=\frac{\Phi_1^\infty+\Phi_i^\infty-2P_\infty\Phi_1^\infty\Phi_i^\infty}{(1-P_\infty\Phi_1^\infty){(1-P_\infty\Phi_i^\infty)}}.
\end{equation}
When $P_\infty\to\infty $, one has
\begin{eqnarray}\label{eq:28}
\Phi_1^\infty\simeq       \frac{1}{P_\infty}\left(1-\frac{1}{P_\infty}+
\frac{2}{P_\infty^2}+                  O(\frac{2}{P_\infty^3})\right),\\
\Phi_i^\infty\simeq       \frac{1}{P_\infty}\left(1+\frac{1}{P_\infty}+
\frac{2}{P_\infty^2}+ O(\frac{2}{P_\infty^3})\right).\label{eq:29}
\end{eqnarray}
Inserting  Eqs.~(\ref{eq:28}) and  (\ref{eq:29})  in Eqs.(\ref{eq:27})
and (\ref{eq:26}) leads to Eq.~(\ref{aex}).

\begin{thebibliography}{99}
\bibitem{VTT93}  This assumption  is  not valid  for  a process  which
evolves far from equilibrium, like the RSA model. In this latter case,
the factorization property is not satisfied. See, P. Viot, G.  Tarjus,
and J. Talbot, Phys. Rev. E {\bf 48}, 480 (1993).
\end{thebibliography}
\end{document}